# Four dimensional quantum oscillator and magnetic-monopole with $U(1)$ dynamical group

Z. Bakhshi$^{a*}$, H. Panahi$^{b}$ and G. Golchehre$^{b}$

$^{a}$Department of Physics, Faculty of Basic Sciences, Shahed University, Tehran 18151/159, Iran.

$^{b}$Department of Physics, University of Guilan, Rasht 51335-1914, Iran.

September 4, 2018

## Abstract

By using of an appropriate transformation, it was shown that the quantum system of 4 dimensional simple harmonic oscillator can describe the motion of a charged particle in the presence of a magnetic monopole field. It was shown that the Dirac magnetic monopole has the hidden algebra of $U(1)$ symmetry and by reducing of the dimensions of space, the $U(1) \times U(1)$ dynamical group for 4D harmonic oscillator quantum system was obtained. Using the group representation and based on explicit solution of the obtained differential equation, the spectrum of system was calculated.

$^{*}$Corresponding author (E-mail: z.bakhshi@shahed.ac.ir)

# 1 Introduction

One of the techniques for calculating energy spectrum eigenvalues and eigenfunctions in the quantum systems is the use of hidden algebra group in the quantum system so that the problem is solvable by the algebra method [1]. Recently, Hurwitze's transformation has been developed for a monopole generation. In this method 2D, 4D, and 8D quantum oscillators can be transformed into the charge-monopole bound systems in $R^2$, $R^3$, and $R^5$, respectively through applying special transformations [2-7]. In this paper, it will be shown that the 4D quantum oscillators can introduce a system with the motion of a charged particle in an interaction of Dirac magnetic-monopole [8,9]. For this purpose, first, based on $KS$ transformation [3], we will present a model for the harmonic oscillator quantum system. Thus, the Four-dimensional space can be transformed to a product of a space with a three Cartesian coordinates and a one-dimensional space [10,11]. Then, for solving obtained differential equations from the three-dimensional spaces in this transformation, two sections for two coordinates systems, namely hyperspherical and parabolic coordinates will be presented in order to separate the variables in the obtained differential equations. It can be easily seen that there is $U(1) \times U(1)$ hidden algebra in this model, so the energy eigenvalues spectrum will be calculated by solving the reduced differential equations based on the hypergeometric functions and Generalized Laguerre polynomials.

# 2 4D quantum oscillator and $KS$ transformation

The Schrödinger equation for the quantum oscillator in the four-dimensional space follows:

$$\sum_{\mu=0}^{3} \frac{\partial^2 \Psi}{\partial u_\mu^2} + \frac{2M}{\hbar^2}\left(E - \frac{M\omega^2 u^2}{2}\right)\Psi = 0, \tag{2.1}$$

where $u^2 = u_\mu u_\mu$. This equation can be considered as a Hamiltonian for a number of four one-dimensional oscillators with equal masses whereas there is no interaction between them. It is clear that the total energy in this system is obtained by the summation of individual energy of particles and the total eigenfunction is also gotten by product of Hermit functions for each particle or each variable $u_\mu$. By applying KS transformation[3]:

$$x_0 = u_0^2 + u_1^2 - u_2^2 - u_3^2$$
$$x_1 = 2(u_0 u_1 + u_2 u_3)$$
$$x_2 = 2(u_0 u_3 - u_1^2 u_2), \qquad (2.2)$$

where

$$R = u^2, \quad r = R^2 = (x_0^2 + x_1^2 + x_2^2)^{1/2}, \qquad (2.3)$$

equation (2.1) is transformed to a equation with the Coulomb potential in the three-dimensional space. By applying the angular transformation, which is presented as the following relation [10]:

$$\gamma = i \ln \frac{u_0 - i u_1}{u_0 + i u_2}, \qquad (2.4)$$

it will be obtained:

$$\frac{1}{2M}\left(i\hbar \frac{\partial}{\partial x_i} - \hbar A_i \widehat{S}\right)^2 \Psi + \frac{\hbar^2}{2Mr^2}\widehat{S}^2 \psi - \frac{e^2}{r}\Psi = \epsilon \Psi, \qquad (2.5)$$

where $i = 0, 1, 2$, $\epsilon = -\frac{M\omega^2}{8}$, $4e^2 = E$, and $\widehat{S} = -i\frac{\partial}{\partial \gamma}$. It is evident, $\widehat{S}$ in equation (2.5) is an extremely small generator of the $U(1)$ algebra [12]. Therefore, the four-dimensional space $R^4$ is represented as a direct product, $R^3 \otimes S^1_{4\pi}$, of the new configuration space $R^3$ with the Cartesian coordinates $x_i \in (-\infty, +\infty)$ and the one-dimensional space with the coordinate $\gamma \in [0, 4\pi)$. In equation (2.5), $A_i$ coordinates are gauge potential of Dirac magnetic-monopole and can be determined using the following relation:

$$\vec{A} = \frac{1}{r(r + x_0)}(-x_2, x_1, 0). \qquad (2.6)$$

Equation (2.5) can be presented as a Dirac magnetic-monopole Hamiltonian for the charge-dyon with the $U(1)$ hidden algebra. It should be mentioned that in the equation (2.1), $\omega$ is considered as a potential constant parameter. So, $E$ is quantized according to $E = \hbar\omega(N+2)$.

Although, in the equation (2.5) $E$ is potential constant parameter and $\omega$ is quantized by $\omega_N = \frac{E}{\hbar(N+2)}$ where $N$ is a natural number in both of equations. Using the generator representation of $U(1)$ group and its application on the wave function as $\Psi = e^{im\gamma}\psi(x_0, x_1, x_2)$, equation (2.5) is converted to the following equation

$$\frac{1}{2M}\left(-i\hbar\frac{\partial}{\partial x_i} - \hbar A_i m\right)^2 \psi + \frac{\hbar^2 m^2}{2Mr^2}\psi - \frac{e^2}{r}\psi = \epsilon\psi. \tag{2.7}$$

In the other words, by applying the representation of group, $\gamma$ coordinate is substitution with the parameter and the integer number $m$ and the one-dimension of space is reduced.

# 3 The solutions of the reduced equation in the hyperspherical coordinates

Let us apply the hyperspherical coordinates in the space $R^4$ according to the following relstions for solving equation (2.7) which describes the charge-dyon system with the $U(1)$-monopole.

$$A_r = A_\theta = 0, \qquad A_\varphi = \frac{1 - \cos\theta}{r\sin\theta}, \tag{3.1}$$

Kinetic energy term of equation (2.7) is gotten as the following form in the hyperspherical coordinates:

$$\frac{1}{2}\left(-i\frac{\partial}{\partial x_i} - A_i m\right)^2 = 1\frac{1}{2}\nabla^2 + \frac{im}{r^2(1+\cos\theta)}\frac{\partial}{\partial\varphi} + \frac{m^2}{2r^2}\left(\frac{1-\cos\theta}{1+\cos\theta}\right). \tag{3.2}$$

By substituting equation (3.2) into the equation (2.7) and considering $i\frac{\partial}{\partial\varphi}$ as an extremely small generator of the $U(1)$ group, it is observed that the $U(1)$ dynamical group is hidden in this system. Therefore the coordinates of the equation can be reduced to $r$ and $\theta$ variables by using of the Lie group representation and assuming the solution of equation (2.7) as:

$$\psi(r, \theta, \varphi) = e^{ip\varphi}R(r)Z(\theta). \tag{3.3}$$

In other words, if a separated constant is considered as $\lambda(\lambda + 1)$, two following ordinary differential equations is gotten as follows:

$$-\frac{1}{2r^2}\frac{\partial}{\partial r}r^2\frac{\partial}{\partial r}R(r) + \left[\frac{1}{2r^2}\lambda(\lambda+1) + \frac{m^2}{2r^2} - \frac{Me^2}{r\hbar^2} - \frac{2M\epsilon}{\hbar^2}\right]R(r) = 0, \qquad (3.4)$$

$$-\frac{1}{\sin\theta}\frac{\partial}{\partial \theta}\sin\theta\frac{\partial}{\partial \theta}Z(\theta) + \left[\frac{p^2}{\sin^2\theta} - \frac{2mp}{1+\cos\theta} + \frac{m^2(1-\cos\theta)}{1+\cos\theta} - \lambda(\lambda+1)\right]Z(\theta) = 0. \qquad (3.5)$$

The equation (3.5) can be solved by assuming $x = \frac{1-\cos\theta}{2}$ and using the transformation $Z(\theta) = F(\theta)M(\theta)$ where $F(\theta)$ is a hypergeometric function. Therefore two un-normalized solutions of the equation (3.5) are gotten based on the hypergeometric function as:

$$Z(\theta) = (\cos\theta)^{m-p-1}(1-\cos\theta)^{p/2} \times_2 F_1(-n_\theta, n_\theta - p + 2m - 1, p + 1; \frac{1-\cos\theta}{2}), \qquad (3.6)$$

and

$$Z(\theta) = (\cos\theta)^{\frac{-2m-p}{2}}(1-\cos\theta)^{p/2} \times_2 F_1(-n_\theta, n_\theta - 2m - 1, p + 1; \frac{1-\cos\theta}{2}). \qquad (3.7)$$

For solving radial term of equation, if the transformation $R(r) = Q(r)L_n^k(r)$ is applied in the equation (3.4) where $L_n^k(r)$ is a generalized Laugerre polynomials, the un-normalized solution of the radial term of this equation can be calculated based on the generalized Laugerre polynomials as follows:

$$R(r) = r^{\frac{k-1}{2}}e^{-\frac{r}{2}}L_n^k(r), \qquad (3.8)$$

where

$$k^2 = 4[m^2 + \lambda(\lambda+1)] + 1, \qquad \frac{2Me^2}{\hbar^2} - \frac{k+1}{2} = n. \qquad (3.9)$$

Also, the energy eigenvalues spectrum in the charge-dyon system with the $U(1)$-monopole in the three-dimensional space can be obtained as:

$$\epsilon = -\frac{Me^4}{2\hbar^2(n+\frac{k+1}{2})^2}, \qquad N = 2n + k + 1. \tag{3.10}$$

Therefore the energy eigenvalues spectrum of four harmonic oscillator can be gotten by substituting parameter definitions of $E$, $N$, and $\omega$ into the relation of the energy eigenvalues (3.10).

## 4 The solutions of the reduced equations in the parabolic coordinates

The parabolic coordinates transformations can be applied to solve the equation (2.7) that is related quantum model. If the three coordinates are considered as the following forms

$$x_1 = \sqrt{\varepsilon\eta}\cos\frac{\varphi}{2}, \qquad x_2 = \sqrt{\varepsilon\eta}\sin\frac{\varphi}{2}, \qquad z = \frac{1}{2}(\eta - \varepsilon). \tag{4.1}$$

$A_i$ coordinates can be determined as $A_\varphi = \frac{2\sqrt{\varepsilon}}{\sqrt{\eta}(\eta+\epsilon)}$ and $A_\eta = A_\varepsilon = 0$ in the parabolic coordinates. By having considered the above new variables, equation (2.7) is rewritten as:

$$-\frac{\hbar^2}{2M}\left(\frac{4}{\varepsilon+\eta}\frac{\partial}{\partial\varepsilon}\varepsilon\frac{\partial}{\partial\varepsilon} - \frac{4}{\varepsilon+\eta}\frac{\partial}{\partial\eta}\eta\frac{\partial}{\partial\eta} - \frac{4}{\varepsilon\eta}\frac{\partial^2}{\partial\varphi^2} + \frac{8mi}{\eta(\varepsilon+\eta)}\frac{\partial}{\partial\varphi}\right)\psi(\varepsilon,\eta,\varphi) +$$
$$\left(\frac{2\hbar^2 m^2}{M}\frac{1}{\eta(\varepsilon+\eta)} - \frac{2e^2}{\varepsilon+\eta} - \varepsilon\right)\psi(\varepsilon,\eta,\varphi) = 0. \tag{4.2}$$

By separating of variables the equation (4.2) is solvable. In the equation (4.2), $i\frac{\partial}{\partial\varphi}$ can be considered as an extremely small generator of the U(1) group as the mentioned it in the previous section. It is clear that $e^{ip\gamma}$ term will be added to the space of representation. Indeed, by assuming the wave function of system as $\psi(\varepsilon,\eta,\varphi) = e^{ip\varphi}N(\varepsilon)Q(\eta)$, equation (4.2) can be reduced to the two following solvable equations:

$$\frac{1}{N}\frac{\partial}{\partial \varepsilon}\varepsilon\frac{\partial N}{\partial \varepsilon} - \frac{p^2}{\varepsilon} + \frac{\epsilon M}{2\hbar^2}\varepsilon = \alpha_1, \qquad (4.3)$$

and

$$\frac{1}{Q}\frac{\partial}{\partial \eta}\eta\frac{\partial Q}{\partial \eta} - \frac{(p-m)^2}{\eta} + \frac{\epsilon M}{2\hbar^2}\eta = \alpha_2, \qquad (4.4)$$

where $\alpha_1$ and $\alpha_2$ are the separated constants which satisfy the relation $\alpha_1 + \alpha_2 = -\frac{e^2 M}{\hbar^2}$ and $p$ is also an integer number. The two equation (4.3) and (4.4) can be transformed to the confluent hypergeometric differential equations. So, the un-normalized wave functions that can satisfy the boundary conditions are represented as:

$$N(\varepsilon) = \varepsilon^p e^{-\frac{\varepsilon}{2}} \times {}_1F_1\left(\frac{2p+1}{2} + \alpha_1, 2p+1; \varepsilon\right), \qquad (4.5)$$

and

$$Q(\eta) = \eta^{(p-m)} e^{-\frac{\eta}{2}} \times {}_1F_1\left(p - m + \alpha_2 + \frac{1}{2}, 2(p-m) + 1; \eta\right). \qquad (4.6)$$

where $p \geq m$ and quantum numbers $p$ and $m$ are integer numbers. On the other hand, the relation $\epsilon = -\frac{M\omega^2}{8}$ due to the condition $\omega = \frac{2\hbar}{M}$ that satisfies the frequency of quantum model. For evanishing divergent solutions in the the confluent hypergeometric functions, parameterized condition is also considered as $c \neq 0, -1, -2, ....$ So, parameter $p$ should be non-negative integer number in the equation (4.4). In addition, $p \geq m$ due to the condition $2p - 2m \neq 0, -1, -2, ...$ in the equation (4.6). Since confluent hypergeometric functions can be changed to a polynomial for $a \neq 0, -1, -2, ...$, the first argument of the confluent hypergeometric functions (4.6) and (4.7) will be written as follows:

$$p + \alpha_1 + \frac{1}{2} = -m_1, \qquad p - m + \alpha_2 + \frac{1}{2} = -m_2. \qquad (4.7)$$

where $m_1$ and $m_2$ are positive integer numbers. If relation (4.7) is substituted in the relation $\alpha_1 + \alpha_2 = -\frac{e^2 M}{\hbar^2}$, the energy eigenvalues will be obtained as following form after

some simplifications:

$$\epsilon = -\frac{Me^2}{2\hbar\hbar^2(2p - m + m_1 + m_2 + 1)^2}. \qquad (4.8)$$

So, the quantum model (2.5) which introduces the motion of a charge particle in the Dirac magnetic-monopole is solvable such as its energy eigenvalue is given by relation (4.8) and its eigenfunction is obtained by $\psi(\gamma, \varepsilon, \eta, \varphi) = e^{im\gamma} e^{ip\varphi} N(\varepsilon) Q(\eta)$.

## 5 Conclusion

When Hamiltonian of the quantum system are written based on the differential generators of Lie algebra by perfect transformations, the energy eigenvalues and the wave eigenfunctions of quantum system are available based on the group representation which has been hidden as a dynamical symmetry in the system. Indeed, the existence of the hidden dynamical group in the non-relativistic quantum systems can be used to determine the energy eigenvalues spectrum and wave eigenfunctions. Actually,the type of group and the dimension of space representation are very important. The dynamical symmetry in Hamiltonian of four dimensional oscillator related to the $U(1)$ dynamical group. So, the part of this system is solvable by applying the generator of this group. Indeed, the one dimension of the representation space reduces by this method. Also, using of hyperspherical and parabolic coordinates can be suitable for separating of variables in the three dimensional spaces of system. So, in addition the energy eigenvalues that are calculated in these coordinates, the wave eigenfunctions of system are gotten based on the hypergeometric functions and generalized Laugerre polynomials.